\documentstyle[twoside,fleqn,espcrc2]{article}

\def\beq{\begin{equation}}
\def\eeq{\end{equation}}
\def\beqa{\begin{eqnarray}}
\def\eeqa{\end{eqnarray}}
\def\lla{\left\langle}
\def\rra{\right\rangle}
\def\za{\alpha}
\def\zb{\beta}

\begin{document}


\pagestyle{empty}

\begin{flushright}
UR-1555\\
ER/40685/924\\
TUIMP-TH-98/102\\
Nov. 1998
\end{flushright}

\vspace*{.2in}

\begin{center}
{\bf \boldmath \protect Could the $\tau$ be 
substantially different from $e$ and $\mu$ in the
supersymmetric standard model ?}$^\star$

\vspace*{.4in}
{\bf Mike Bisset$^*$,
Otto C. W. Kong$^{\dagger}$, Cosmin Macesanu$^{\dagger}$, 
and Lynne H. Orr$^{\dagger}$}

\vspace*{.4in}
{\it $^*$Department of Physics, \\ Tsinghua University,  Beijing 100084, China\\[.2in]
$^{\dagger}$ Department of Physics and Astronomy,\\
University of Rochester, Rochester NY 14627-0171.\\
}

\vspace*{.8in}
{Abstract}
\end{center}
R-parity stands as an {\it ad hoc} assumption in the most popular version 
of the supersymmetric standard model. More than fifteen years' studies of 
R-parity violations have been restricted to various limiting
scenarios. We illustrate how the single-VEV parametrization
provides a workable framework to analyze the  phenomenology
of the complete theory of supersymmetry without R-parity. In our 
comprehensive study of various aspects of the resulting 
leptonic phenomenology at tree-level, we find that the physical  
$\tau$ lepton could actually bear  substantial gaugino and higgsino
components, making it very different from the $e$ and the $\mu$.

\noindent

\vfill
\noindent --------------------- \\
$^\star$ Talk given by O.K. at Tau 98 conference, 
Santander, Spain, Sep 1998 --- submission for proceedings.  

\clearpage


\pagestyle{plain}
\addtocounter{page}{-1}

\title{Could the $\tau$ be substantially different from 
$e$ and $\mu$ in the supersymmetric standard model ?
} 
       
\author{Mike Bisset\address{Department of Physics, Tsinghua University,  Beijing 100084, China}, and
Otto C. W. Kong$^b$
\thanks{Participant of the conference who presented the talk.}
, Cosmin Macesanu$^b$, 
and Lynne H. Orr\address{Department of Physics and Astronomy,
University of Rochester, Rochester NY 14627-0171, U.S.A.
}}
             

\begin{abstract}
R-parity stands as an {\it ad hoc} assumption in the most popular version 
of the supersymmetric standard model. More than fifteen years' studies of 
R-parity violations have been restricted to various limiting
scenarios. We illustrate how the single-VEV parametrization
provides a workable framework to analyze the  phenomenology
of the complete theory of supersymmetry without R-parity. In our 
comprehensive study of various aspects of the resulting 
leptonic phenomenology at tree-level, we find that the physical  
$\tau$ lepton could actually bear  substantial gaugino and higgsino
components, making it very different from the $e$ and the $\mu$.
\end{abstract}
\maketitle

\section{INTRODUCTION}
We are going to illustrate how  the $\tau$ could be substantially different 
from $e$ and $\mu$ in the supersymmetric standard model. In a
supersymmetric version of the standard model (SM), the leptons
could assume a new identity unless an {\it ad hoc} symmetry called
R-parity is imposed. We will show that R-parity is not theoretically
well motivated, and the present experimental constraints from leptonic
phenomenology allow significant violation of R-parity. In particular, the 
$\tau$ might contain substantial components in the gaugino and higgsino
directions, making it very different from the $e$ and the $\mu$.

\section{SUPERSYMMETRY AND R-PARITY}
Supersymmetry is a symmetry between bosons and fermions.
To produce a supersymmetric version of the SM, one has to be able to
put its content in terms of superfields, which contain both bosonic
and fermionic components. Each matter field is to be embedded
into a chiral superfield containing a complex scalar and a Weyl fermion.
It is then obvious that we need the $\hat{Q}_i$, $\hat{U}_i^c$,
$\hat{D}_i^c$, and $\hat{E}_i^c$ superfields with three flavors,
{\it i.e.} $i=1$ to $3$. Due to the  holomorphic properties of the
superpotential, separate superfields are needed to provide the Higgs
scalars for the mass generation of the up- and down-sector quarks.
For the former, a $\hat{H}_u$ is introduced. For the latter, a
$\hat{H}_d$ as a superfield will be identical to a $\hat{L}$ from
a leptonic doublet. However, the fermionic component of $\hat{H}_u$,
the higgsino, will spoil the beautiful gauge anomaly cancellation of
the SM fermion spectrum, unless a conjugate fermionic
doublet is also added. Hence, four superfields (flavors) with the quantum
number of $L$ or $H_d$ are needed. We denote them by $\hat{L}_\za$,
$\za=0$ to $3$.

R-parity is defined by
\[
{\cal R} = (-1)^{3B+L+2S}
\]
where $B, L,$ and $S$ are the baryon number, the lepton number,
and the spin of a superfield component respectively. As superfields,
$\hat{Q}_i\;,  \hat{U}_i^c\;, \hat{D}_i^c\;, \hat{E}_i^c$ and
$\hat{L}_i$, taking as those three from the leptonic doublets, are
odd, while $\hat{H}_u$ and $ \hat{H}_d$ are even under R-parity.
Componentwise, all SM particles have even R-parity while all
the superpartners have odd R-parity. Hence,  $ \hat{H}_d$ and 
$\hat{L}_i$ no longer have the same quantum number. So, adding
supersymmetry and  R-parity  is to first put in a symmetry between
fermions and bosons and then put in another symmetry to distinguish
the known fermions from the fermionic partners of the scalars,
that we believe have to be there. 

The most general superpotential admissible by the SM gauge 
symmetries can be written as 
\beqa \small
W \!\!\!\!&=&\!\!\!\! \varepsilon_{ab}\big[ \mu_{\za}  
\hat{L}_{\za}^a \hat{H}_u^b
+ h_{ik}^u \hat{Q}_i^a   \hat{H}_{u}^b \hat{U}_k^{\scriptscriptstyle C}
+ \lambda_{i\za k}^{'} \hat{Q}_i^a \hat{L}_{\za}^b 
\hat{D}_k^{\scriptscriptstyle C}   \nonumber\\
&&
+ \lambda_{\za \zb k}  \hat{L}_{\za}^a  
 \hat{L}_{\zb}^b \hat{E}_k^{\scriptscriptstyle C}
\big] + \lambda_{i jk}^{''}  \hat{D}_i^{\scriptscriptstyle C}  
\hat{D}_j^{\scriptscriptstyle C}  \hat{U}_k^{\scriptscriptstyle C}\; ,
\eeqa \normalsize
where  $(a,b)$ are $SU(2)$ indices, $(i,j,k)$ are family (flavor) indices,
$(\za,\zb)$ are (extended) flavor indices from $0$ to $3$,
and  $\hat{L}_{\za}$'s denote the four doublet superfields with $Y=-1/2$.  
$\lambda$ and $\lambda^{''}$ are antisymmetric in the first two indices as
required by  $SU(2)$ and  $SU(3)$ product rules respectively.
This is shown expicitly here only for $SU(2)$, with
$\varepsilon = \left({\begin{array}{cc} 0 & -1 \\ 1 &  0 \end{array}}\right)$.
The  $SU(3)$ indices are  suppressed.
In the limit where $\lambda_{ijk}\;, \lambda^{'}_{ijk}\,, \lambda^{''}_{ijk}$
and $\mu_{i}$  all vanish, one recovers the R-parity conservating
result (the minimal supersymmetric SM), with the following matching to 
the common notation of the latter:
\beqa
&\bullet& \hat{L}_0 \longrightarrow \hat{H}_d \nonumber \\
&\bullet& \lambda^{'}_{i0k} \longrightarrow h^{d}_{ik} \nonumber \\
&\bullet& \lambda_{i0k} -\lambda_{0ik}  \longrightarrow h^{e}_{ik} 
\nonumber\;. \eeqa

The superpotential contains both B- and L-violating terms resulting in 
tree-level superparticle mediated proton decay.
The experimental limit on the
proton life-time then requires the product of the relevant couplings
($\lambda^{'}$ and $\lambda^{''}$)
to be of the order $10^{-27}$\cite{pdecay}. While the only natural
way to satisfy the stringent proton decay constraint is to forbid the 
process by a symmetry, R-parity is not the only option, nor is it
necessarily the best one. For example, a baryon-parity enforcing 
$\lambda^{''}=0$ or  other discrete symmetries can be
used\cite{bp}. The former option has  an  advantage over R-parity --- 
it forbids, in addition to the dimension-4 B-violation terms ($\lambda^{''}$), 
dangerous terms of dimension 5. These alternatives
are usually much less restrictive; they allow quite a number of 
R-parity violating (RPV) terms in the Lagrangian for which there are
interesting experimental constraints\cite{rpv}.

We consider it more interesting 
to adopt a purely phenomenological approach to study of the complete
supersymmetric standard model without R-parity and analyze how 
the overall parameter space is restricted by the various constraints,
as well as looking for potential experimental signals of R-parity
violation.

\section{THE SINGLE-VEV PARAMETRIZATION}
The above suggestion sounds naively straight forward; however, its
implementation demands a careful consideration. The large number
of extra parameters involved makes the task difficult to manage.
For instance, the tree-level color-single charged fermion mass
matrix, in a generic flavor basis, is given by
\footnotesize
\beq 
{\cal{M}_{\scriptscriptstyle C}} =
 \left(
{\begin{array}{ccccc}
{M_{\scriptscriptstyle 2}} &  
\frac{g_{\scriptscriptstyle 2}{v}_{\scriptscriptstyle u}}{\sqrt 2} 
 & 0  \\
 \frac{g_{\scriptscriptstyle 2}{v}_{\scriptscriptstyle d}}{\sqrt 2} &
 {{ \mu}_{\scriptscriptstyle 0}} &   
 -h^e_{ik}  \frac{{v}_{\scriptscriptstyle i}}{\sqrt 2} \\ 
 \frac{g_{\scriptscriptstyle 2}{v}_{\scriptscriptstyle j}}{\sqrt 2} 
& {{ \mu}_{j}} &
 h^e_{jk}  \frac{{v}_{\scriptscriptstyle d}}{\sqrt 2} 
+ 2\lambda_{jik}\frac{{v}_{\scriptscriptstyle i}}{\sqrt 2}\\
\end{array}}
 \right) \; ,
\eeq \normalsize
where we have suppressed the last three rows and columns with indices
$j$ and $k$ going from 1 to 3. Note the summation over $i$; and that
we have written the matrix in such a way as to distinguish the R-parity
conservating and violating terms. The VEV's are given by
\beqa
\frac{{v}_{\scriptscriptstyle u}}{\sqrt 2}= \lla \hat{H}_u \rra \; , \hspace*{.2in}
\frac{{v}_{\scriptscriptstyle i}}{\sqrt 2}= \lla \hat{L}_i \rra \; ,  \nonumber \\
\hbox{and}\hspace*{.2in} \frac{{v}_{\scriptscriptstyle d}}{\sqrt 2}\equiv
\frac{{v}_{\scriptscriptstyle 0}}{\sqrt 2}= \lla \hat{L}_0 \rra  \; . \nonumber
\eeqa
Remember the only knowledge we have about the matrix entries is that of the 
gauge couplings $g_{\scriptscriptstyle 1}$ and $g_{\scriptscriptstyle 2}$,
the overall magnitude of the electroweak symmetry breaking VEV's
\beq
|{v}_{\scriptscriptstyle u}|^2 + \sum |{v}_{\scriptscriptstyle \za}|^2 = v^2
=246 \hbox{GeV} \; ,
\eeq
and that they have to yield the correct mass eigenvalues for the physical
leptons ($\ell$), namely $m_e$, $m_\mu$, and $m_\tau$. However, 
we need to have a good knowledge of the real nature of the $\ell$'s 
before we can  study the experimental signatures of the heavier
particles.

The problem is solved in the single-VEV parametrization, which is
nothing more than writing the Lagrangian in the most convenient set of flavor bases.
More details of the parametrization are discussed in Ref\cite{I}. Under
the framework, the ${v}_{\scriptscriptstyle i}$'s and the off diagonal
$h^e_{jk}$'s are set to zero. This is an optimal exploitation of the
freedom in the choice of flavor bases. The above mass matrix is then 
simplified to
\beq
\footnotesize 
{\mathcal{M}_{\scriptscriptstyle C}} =
 \left(
{\begin{array}{ccccc}
{M_{\scriptscriptstyle 2}} &  
\frac{g_{\scriptscriptstyle 2}{v}_{\scriptscriptstyle u}}{\sqrt 2}  
& 0 & 0 & 0 \\
 \frac{g_{\scriptscriptstyle 2}{v}_{\scriptscriptstyle d}}{\sqrt 2} & 
 {{ \mu}_{\scriptscriptstyle 0}} & 0 & 0 & 0 \\
0 &  {{ \mu}_{\scriptscriptstyle 1}} & {{m}_{\scriptscriptstyle 1}} & 0 & 0 \\
0 & {{ \mu}_{\scriptscriptstyle 2}} & 0 & {{m}_{\scriptscriptstyle 2}} & 0 \\
0 & {{ \mu}_{\scriptscriptstyle 3}} & 0 & 0 & {{m}_{\scriptscriptstyle 3}}
\end{array}}
\right)  \; .
\eeq \normalsize
Here, each $m_i$ is a physical leptonic mass in the limit where the
corresponding $\mu_i$ is zero. In the general case, the correct value
of the $m_i$'s can be determined, at least numerically, to guarantee
acceptable eigenvalues of the $\ell_i$ masses. Hence, the mass eigenstates
and their exact nature, such as their gaugino and higgsino contents,
can be found. In particular, our result shows that ${ \mu}_{\scriptscriptstyle 3}$,
which characterizes the gaugino and higgsino contents of the $\tau$, is
not necessarily small.
 The  chargino masses now also depend on  $\mu_i$'s,
with interesting implications. An example is illustrated in Fig.1.

\vspace{2.35in}
\includegraphics{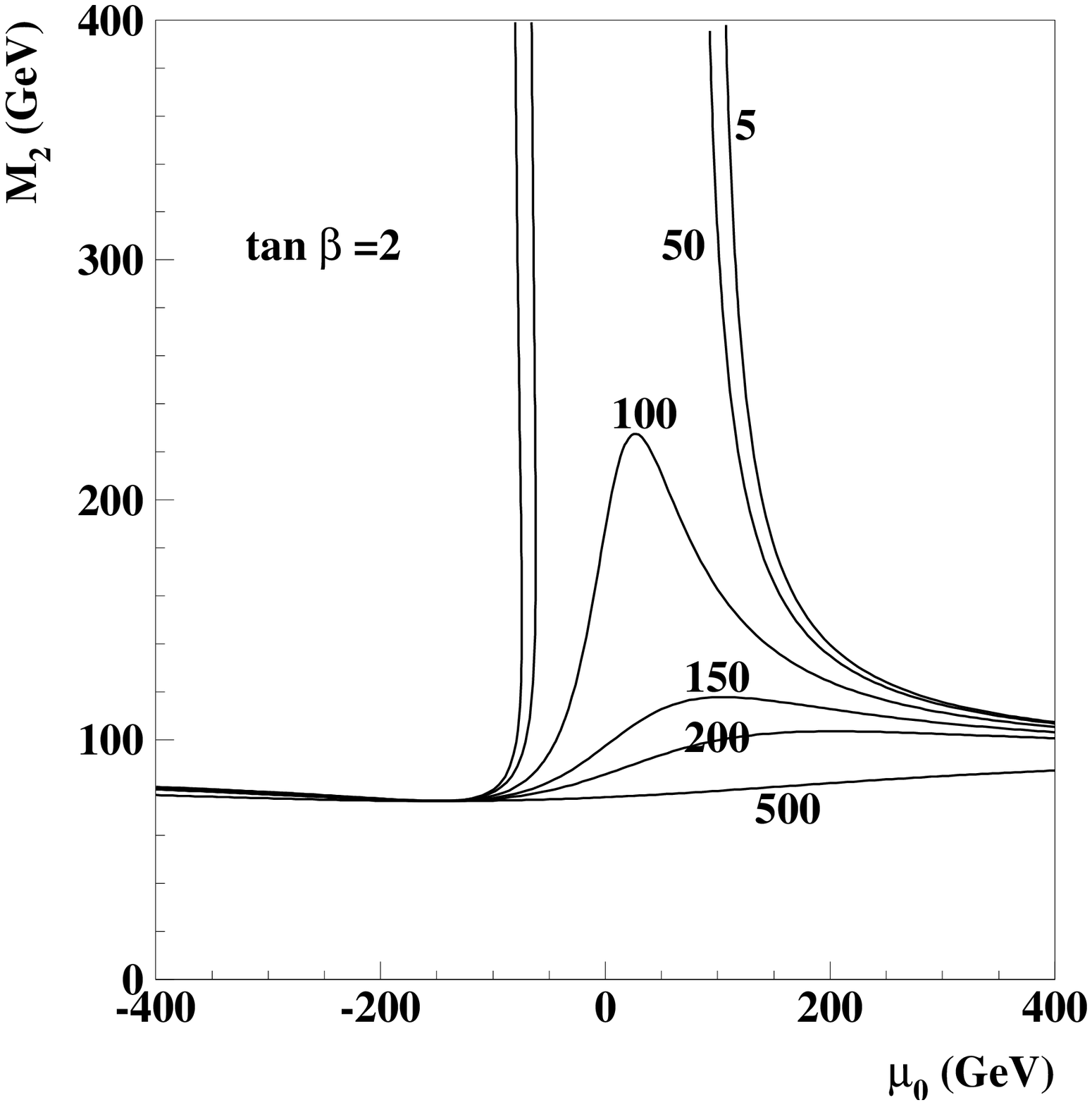}
{Figure 1.
Minimum values of ${\mu}_{\scriptscriptstyle 5}$ (in GeV) required to give
both chargino masses above $90\, \hbox{GeV}$.
($\mu_{\scriptscriptstyle 5} = \sqrt{\mu_{\scriptscriptstyle 1}^2 
+\mu_{\scriptscriptstyle 2}^2 +\mu_{\scriptscriptstyle 3}^2}$)
}

\section{A NEUTRINO MASS}
Similarly, a simple form of the neutral fermion mass matrix is obtained:
\beq 
\tiny
\label{mn}
{\cal{M}_{\scriptscriptstyle N}} 	
\!\!=\!\!  \left(\!\!\!\!
{\begin{array}{ccccccc}
{{M}_{\scriptscriptstyle 1}} \!\!\!&\!\!\! 0 \!\!\!&\!\!\!  \frac {{g}_{1}{v}_{u}}{2}
 \!\!\!&\!\!\!  -\frac{{g}_{1}{v}_{d}}{2} \!\!\!&\!\!\! 0 \!\!\!&\!\!\! 0 \!\!\!&\!\!\! 0 \\
0 \!\!\!&\!\!\! {{M}_{\scriptscriptstyle 2}} \!\!\!&\!\!\!  -\frac{{g}_{2}{v}_u}{2} \!\!\!&\!\!\! 
\frac{{g}_{2}{v}_{d}}{2} \!\!\!&\!\!\! 0 \!\!\!&\!\!\! 0 \!\!\!&\!\!\! 0 \\
 \frac {{g}_{1}{v}_{u}}{2} \!\!\!&\!\!\!   -\frac{{g}_{2}{v}_u}{2} \!\!\!&\!\!\! 0 \!\!\!&\!\!\! 
 - {{\mu}_{\scriptscriptstyle 0}} \!\!\!&\!\!\!  - {{ \mu}_{\scriptscriptstyle 1}} 
 \!\!\!&\!\!\!  - {{ \mu}_{\scriptscriptstyle 2}} \!\!\!&\!\!\!  - {{ \mu}_{\scriptscriptstyle 3}} \\
 -\frac{{g}_{1}{v}_{d}}{2} \!\!\!&\!\!\! \frac{g_{2}{v}_d}{2}
 \!\!\!&\!\!\!  - {{ \mu}_{0}} \!\!\!&\!\!\! 0 \!\!\!&\!\!\! 0 \!\!\!&\!\!\! 0 \!\!\!&\!\!\! 0 \\ 
0 \!\!\!&\!\!\! 0 \!\!\!&\!\!\!  - {{ \mu}_{\scriptscriptstyle 1}} \!\!\!&\!\!\! 0 \!\!\!&\!\!\! 0 \!\!\!&\!\!\! 0 \!\!\!&\!\!\! 0 \\
0 \!\!\!&\!\!\! 0 \!\!\!&\!\!\!  - {{ \mu}_{\scriptscriptstyle 2}} \!\!\!&\!\!\! 0 \!\!\!&\!\!\! 0 \!\!\!&\!\!\! 0 \!\!\!&\!\!\! 0 \\
0 \!\!\!&\!\!\! 0 \!\!\!&\!\!\!  - {{ \mu}_{\scriptscriptstyle 3}} \!\!\!&\!\!\! 0 \!\!\!&\!\!\! 0 \!\!\!&\!\!\! 0 \!\!\!&\!\!\! 0
\end{array}}
\!\!\!\! \right) \; .
\eeq \normalsize
 
Two neutrino eigenstates are left massless at the tree level, while the 
third one gains a mass through the RPV-couplings ($\mu_i$'s) to the higgsino.
 In fact one  can use a simple 
rotation to decouple the massless states. The remaining $5\times 5$ mass 
matrix is then given by
\beq \footnotesize \label{mn5}
{\cal{M}_{\scriptscriptstyle N}}^{\!\!\scriptscriptstyle (5)} \! 	
=  \! \left(\!\!\!
{\begin{array}{ccccc}
{M}_{\scriptscriptstyle 1} \!\!&\!\! 0 \!\!&\!\!  \frac {{g}_{1}{v}_{u}}{2}
 \!\!&\!\!  -\frac{{g}_{1}{v}_{d}}{2} \!\!&\!\! 0 \\
0 \!\!&\!\! {{M}_2} \!\!&\!\!  -\frac{{g}_{2}{v}_u}{2} 
\!\!&\!\! \frac{{g}_{2}{v}_{d}}{2} \!\!&\!\! 0 \\
 \frac {{g}_{1}{v}_{u}}{2} \!\!&\!\!   -\frac{{g}_{2}{v}_u}{2} \!\!&\!\! 0 \!\!&\!\!  
 - {{\mu}_{\scriptscriptstyle 0}} \!\!&\!\!  - {{\mu}_{\scriptscriptstyle 5}}  \\
 -\frac{{g}_{1}{v}_{d}}{2} \!\!&\!\! \frac{g_{2}{v}_d}{2}
 \!\!&\!\!  - {{\mu}_{\scriptscriptstyle 0}} \!\!&\!\! 0 \!\!&\!\! 0  \\ 
0 \!\!&\!\! 0 \!\!&\!\!  - {{\mu}_{\scriptscriptstyle 5}} \!\!&\!\! 0 \!\!&\!\! 0 
\end{array}} \!\!\!\right)  ,
\eeq \normalsize
where 
\beq
\mu_{\scriptscriptstyle 5} = \sqrt{\mu_{\scriptscriptstyle 1}^2 
+\mu_{\scriptscriptstyle 2}^2 +\mu_{\scriptscriptstyle 3}^2} \; ;
\eeq
and the corresponding massive neutrino state is given,
in terms of the original neutrino basis,  by
\beq
\left|\nu_{\scriptscriptstyle 5}\rra = 
\frac{\mu_{\scriptscriptstyle 1}}{\mu_{\scriptscriptstyle 5}}
\left|\psi^{\scriptscriptstyle 1}_{\scriptscriptstyle L_1}\rra
+ \frac{\mu_{\scriptscriptstyle 2}}{\mu_{\scriptscriptstyle 5}}
\left|\psi^{\scriptscriptstyle 1}_{\scriptscriptstyle L_2}\rra
+  \frac{\mu_{\scriptscriptstyle 3}}{\mu_{\scriptscriptstyle 5}}
\left|\psi^{\scriptscriptstyle 1}_{\scriptscriptstyle L_3}\rra  \; ,
\eeq
a mixture of all three of them in 
general. Adopting a $``$seesaw" approximation gives the neutrino mass 
\beq
|m_{\nu_{\scriptscriptstyle 5}}|  = 
\frac{ {\mu}_{5}^{2} {v}^{2} \cos^2\!\!\zb 
\left( x{g}_{\scriptscriptstyle 2}^{2} 
+ {g}_{\scriptscriptstyle 1}^{2} \right) }
{\mu_{\scriptscriptstyle 0} \left[ 4xM_{\scriptscriptstyle 2}
 \mu_{\scriptscriptstyle 0} -
 \left( x{g}_{\scriptscriptstyle 2}^{2}+{g}_{1\scriptscriptstyle }^{2}\right) 
{v}^2 \sin\!2{\zb} \right] } , 
\eeq
where we have substituted $v_{\scriptscriptstyle d}=v\cos\!{\zb}$, 
$v_{\scriptscriptstyle u}=v\sin\!{\zb}$,
and $M_{\scriptscriptstyle 1}=xM_{\scriptscriptstyle 2}$. Note that for large 
$\tan\!{\zb}$, $\cos\!{\zb}$ is a strong suppression factor. In order to
look at potential large  $\mu_{\scriptscriptstyle 5}$ values, we can perform
an alternative perturbative analysis,  diagonalizing exactly the matrix without  
the EW-symmetry breaking terms, only to put them back as perturbation.
This yields
\beq \label{mnu}
|m_{\nu_{\scriptscriptstyle 5}}|=  \frac{1}{4} 
\frac{{\mu}_{\scriptscriptstyle 5}^{2} {v}^{2} \cos\!\!^2\zb 
\left(x{g}_{\scriptscriptstyle 2}^{2} + {g}_{\scriptscriptstyle 1}^{2} \right)}
{ \left( \mu_{\scriptscriptstyle 0}^{2} + \mu_{\scriptscriptstyle 5}^{2} 
\right) xM_{\scriptscriptstyle 2} }\; ,
\eeq
giving
\beq \label{mu5}
\mu_{\scriptscriptstyle 5}^2
< \frac {4 {x}{\mu}_{\scriptscriptstyle 0}^{2}M_{\scriptscriptstyle 2} 
m_{\nu_{\scriptscriptstyle 5}}^b}{{v}^{2} \cos^{2}\!\!{\zb}
\left( x{g}_{\scriptscriptstyle 2}^{2} + {g}_{\scriptscriptstyle 1}^{2} \right) 
- 4{x}M_{\scriptscriptstyle 2}  
m_{\nu_{\scriptscriptstyle 5}}^b} \; ,
\eeq
for an experimental neutrino mass bound $m_{\nu_{\scriptscriptstyle 5}}^b$.
As $M_{\scriptscriptstyle 2}$ increases, the denominator above drops to zero,
beyond which there is {\it no}  bound on $\mu_{\scriptscriptstyle 5}$. 

\addtocounter{figure}{1}
\begin{figure*}[t]
\vspace{3in}
\includegraphics{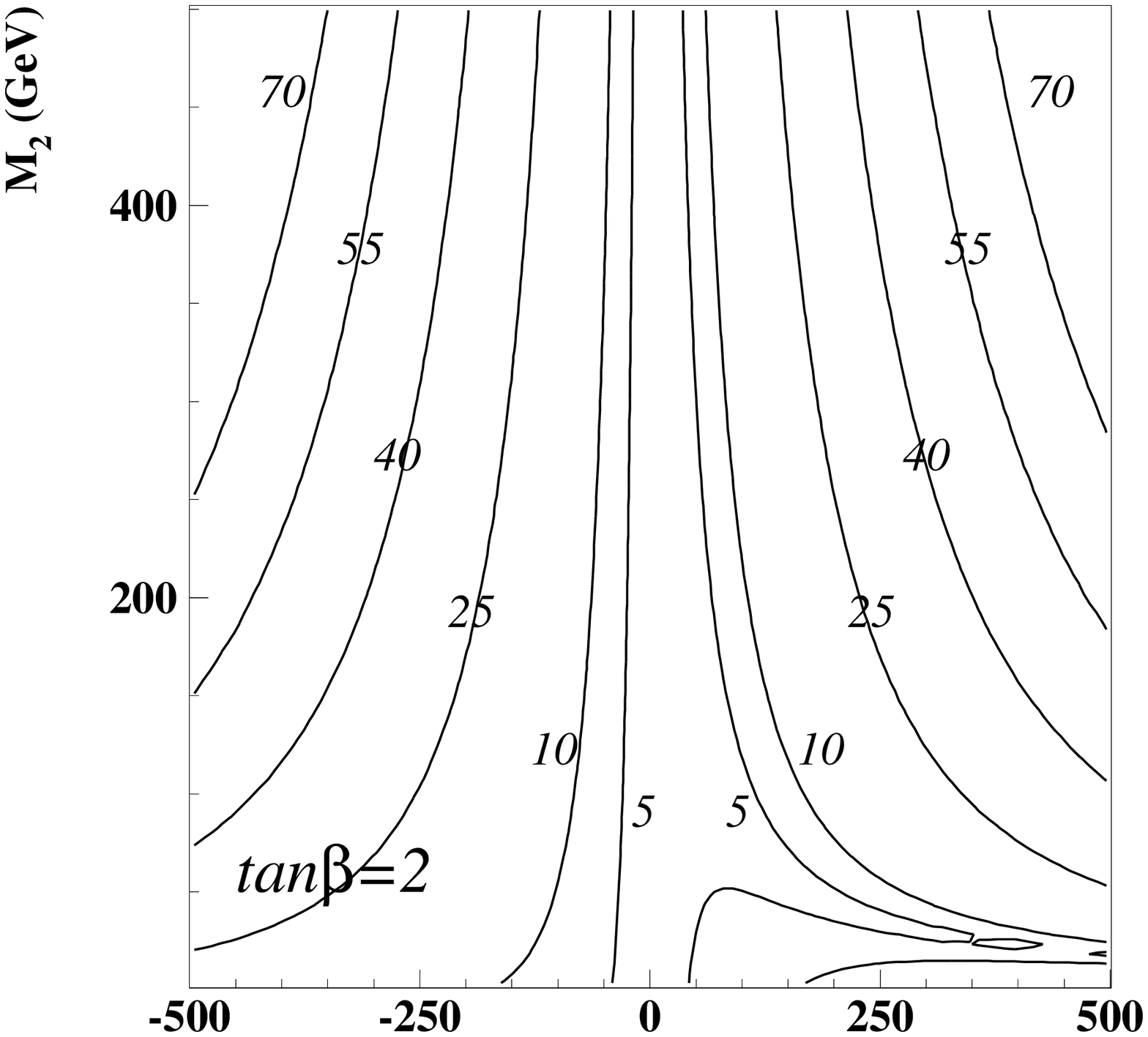}
\includegraphics{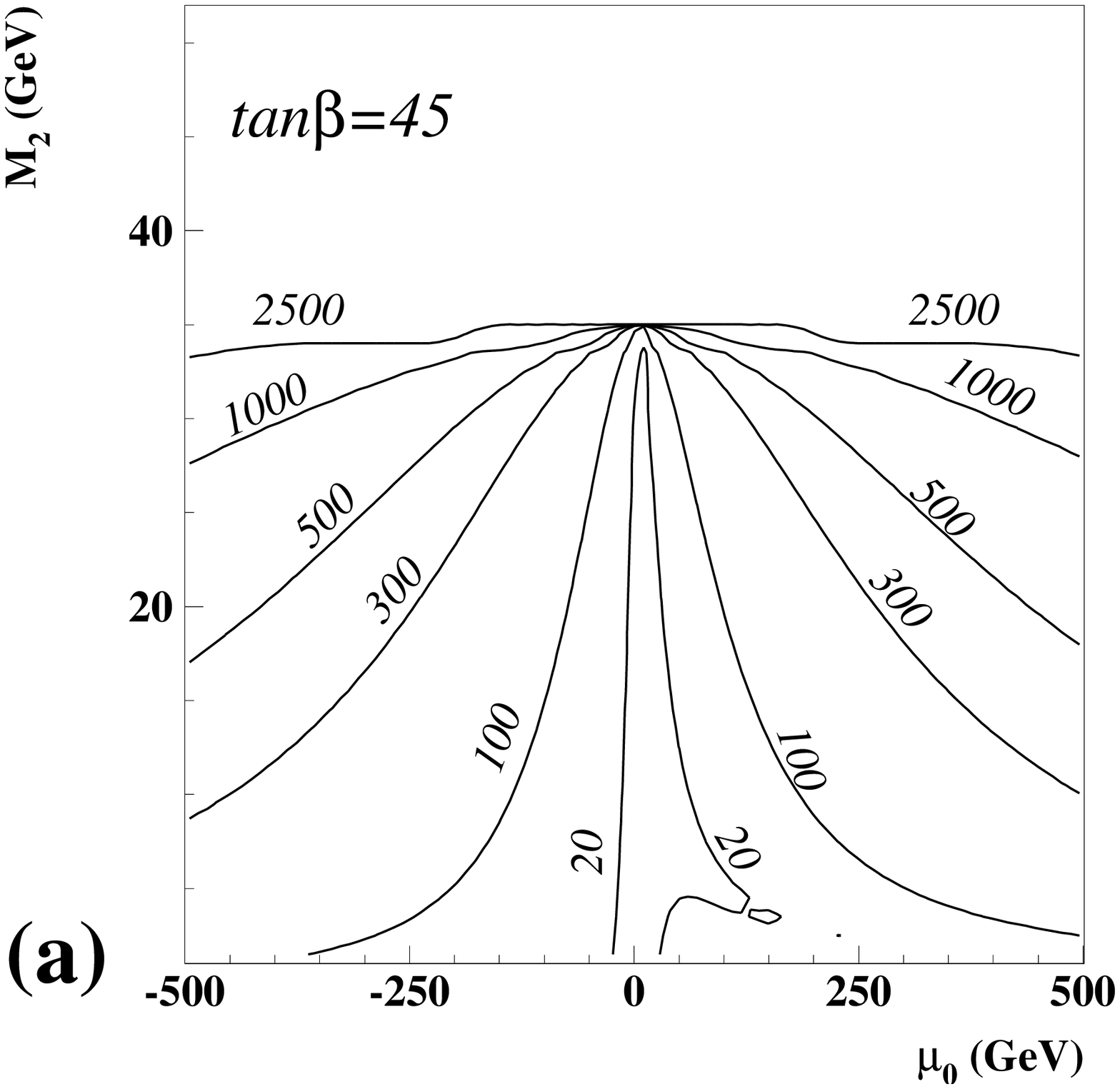}
\includegraphics{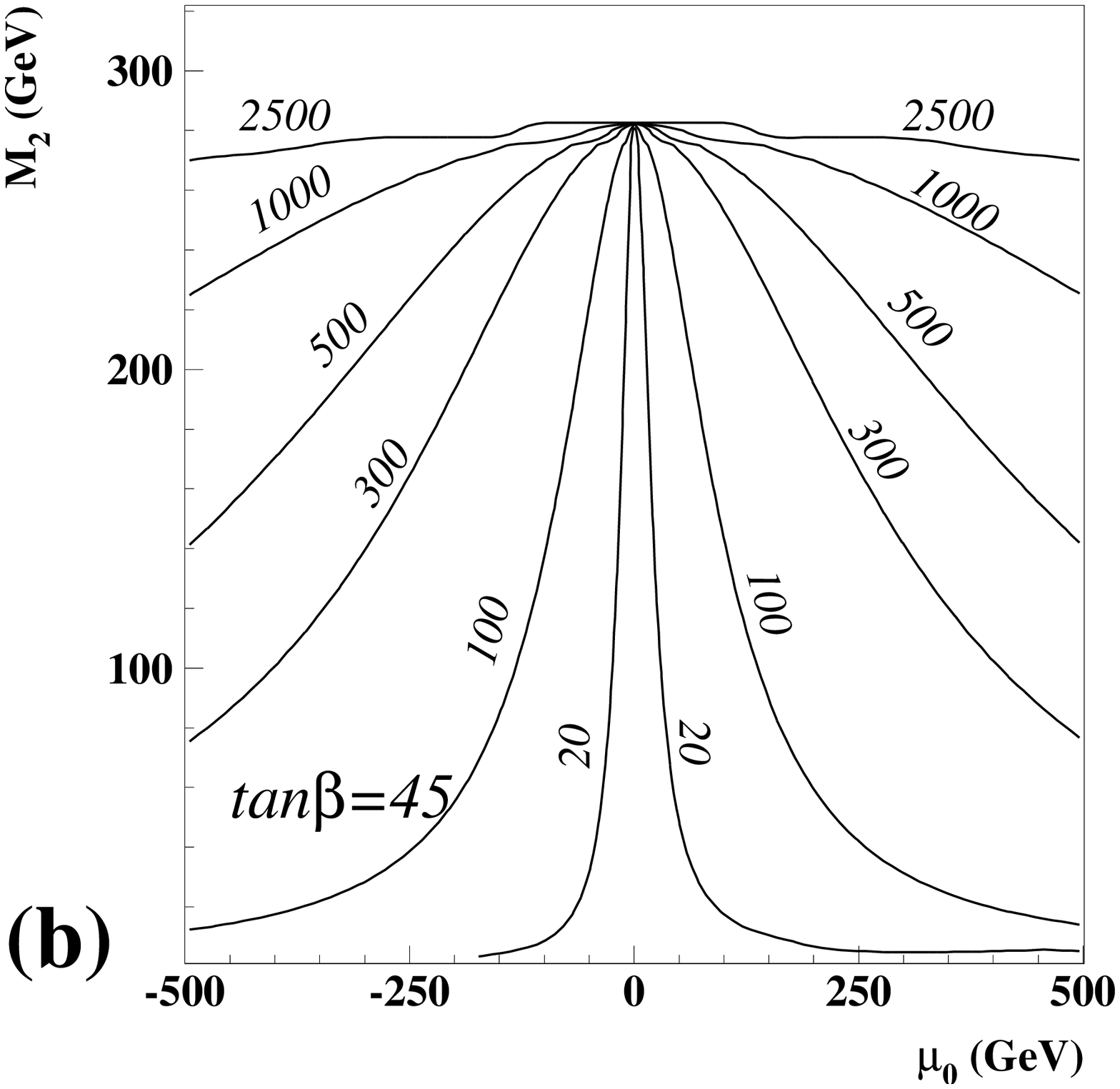}
\includegraphics{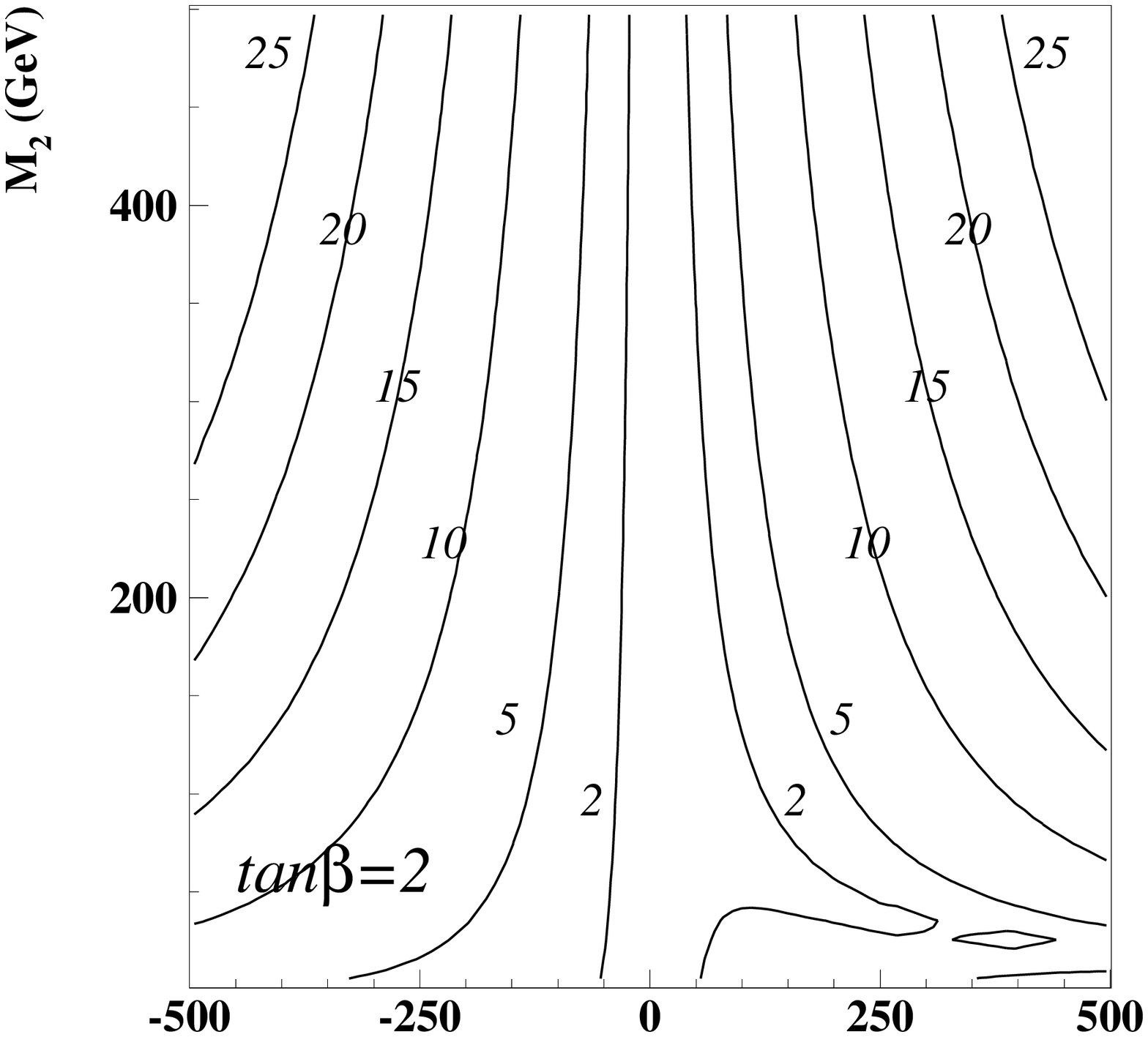}
\caption{Maximum allowed values of ${\mu}_{\scriptscriptstyle 5}$ (in GeV) consistent 
with a specific neutrino mass bound: (a) the absolute bound 
$m_{{\nu}_{\scriptscriptstyle 3}} < 149\, \hbox{MeV}$; (b) machine bound
$m_{{\nu}_{\tau}} < 18.2\, \hbox{MeV}$, applicable for
${\mu}_{\scriptscriptstyle 1}:{\mu}_{\scriptscriptstyle 2}:
{\mu}_{\scriptscriptstyle 3} = 0:0:1$.
The region above or outside of a given contour 
is excluded for ${\mu}_{\scriptscriptstyle 5}$'s above the indicated value.
}
\end{figure*} 

The perturbative result in Eq.(\ref{mu5}) is borne out by exact numerical 
results from diagonalizing the neutral fermion mass matrix, as illustrated in 
Fig. 2, where the machine $\nu_\tau$ bound of $18.2\,\mbox{MeV}$\cite{182} and the generic neutrino mass bound of $149\,\mbox{MeV}$, from charged
current data\cite{BFM}. The dependence of the 
${\mu}_{\scriptscriptstyle 5}$ bounds on $\tan\!\zb$ is striking.  For $\tan\!{\zb}=45$, values of ${\mu}_{\scriptscriptstyle 5}$ 
in the hundreds of GeV or beyond are not ruled out.

There are potentially much stronger bounds on  neutrino masses from
cosmological considerations which however depend on the decay modes
and other assumptions so that a neutrino mass above an MeV 
is not definitely ruled out. For much smaller neutrino masses, such
as those fitting naturally fitting the oscillation scenario indicated by recent
results from Super-Kamiokande\cite{nuK}, the $\mu_i$'s would have to be
below the MeV scale unless much larger $\tan\!\zb$ value is assumed.

\section{LEPTONIC PHENOMENOLOGY AND CONSTRAINTS}
With the fermion mass matrices discussed above, we can elicit the properties
of the mass eigenstates and their interactions in any region of the parameter
space. The matrices are simple enough that effective perturbative 
approximations can be applied in the study of the leptons $\ell$'s and $\nu$'s,
and exact numerical diagonalizations of both mass matrices can be
performed. A rich implication in leptonic phenomenology results from 
nonzero $\mu_i$'s, which can also be used to place more experimental
constraints on the admissible values of the latter.

\begin{table*}
\setlength{\tabcolsep}{1.5pc}
\caption{Summary of phenomenological constraints incorporated into our study: details of notations and sources of experimental bounds can be
found in our paper\cite{II}.}
 \normalsize
\begin{tabular*}{\textwidth}{@{}l@{\extracolsep{\fill}}cc}
\hline \hline 
       {\,\,\,\, Quantity } & 
	{${\mu}_{i}$  combo. constrained }&
       { Experimental bounds } \\ \hline
       & & \\[-.1in]
       \framebox{$Z^0$-coupling:} & & \\[.1in]
      $\bullet$ $Br$($\mu^- \to e^- e^+e^-$)
       &  $|\mu_{\scriptscriptstyle 1}\mu_{\scriptscriptstyle 2}|$ 
       &  $<1.0\times 10^{-12}$ \\[0in]
        $\bullet$ $Br$($\tau^- \to e^- e^+e^-$)
       & $|\mu_{\scriptscriptstyle 1}\mu_{\scriptscriptstyle 3}|$ 
       & $<2.9\times 10^{-6}$\\[0in]
        $\bullet$ $Br$($\tau^{-} \to \mu^{-} e^+ e^-$)
       & $|\mu_{\scriptscriptstyle 2}\mu_{\scriptscriptstyle 3}|$ 
       &  $<1.7\times 10^{-6}$\\[0in]
       $\bullet$ $Br$($\tau^{-} \to \mu^{+} e^- e^-$)
       &  $|\mu_{\scriptscriptstyle 1}^2\mu_{\scriptscriptstyle 2}\mu_{\scriptscriptstyle 3}|$ 
       & $<1.5\times 10^{-6}$\\[0in]
       $\bullet$ $Br$($\tau^{-} \to e^{-} \mu^+ \mu^-$)
       &  $|\mu_{\scriptscriptstyle 1}\mu_{\scriptscriptstyle 3}|$ 
       & $<1.8\times 10^{-6}$\\[0in]
       $\bullet$ $Br$($\tau^{-} \to e^{+} \mu^- \mu^-$)
       &  $|\mu_{\scriptscriptstyle 1}\mu_{\scriptscriptstyle 2}^2\mu_{\scriptscriptstyle 3}|$ 
       & $<1.5\times 10^{-6}$\\[0in]
       $\bullet$ $Br$($\tau^- \to \mu^- \mu^+ \mu^-$)
       &  $|\mu_{\scriptscriptstyle 2}\mu_{\scriptscriptstyle 3}|$ 
       & $<1.9\times 10^{-6}$\\[0in]
       $\bullet$ $Br$($Z^0 \to e^{\pm} \mu^{\mp}$)
       &  $|\mu_{\scriptscriptstyle 1}\mu_{\scriptscriptstyle 2}|$ 
       &  $<1.7\times 10^{-6}$ \\[0in]
       $\bullet$ $Br$($Z^0 \to e^{\pm} \tau^{\mp}$)
       &  $|\mu_{\scriptscriptstyle 1}\mu_{\scriptscriptstyle 3}|$ 
       &  $<9.8\times 10^{-6}$ \\[0in]
       $\bullet$ $Br$($Z^0 \to \mu^{\pm} \tau^{\mp}$)
       &  $|\mu_{\scriptscriptstyle 2}\mu_{\scriptscriptstyle 3}|$ 
       &  $<1.2\times 10^{-5}$ \\[0in]
       $\bullet$ $Br$($Z^0 \to \chi^{\pm} \ell^{\mp}$)
       &  {complicated} 
       &  $< 1.0\times 10^{-5}$ \\[0in]
       $\bullet$ $Br$($Z^0 \to \chi^{\pm} \chi^{\mp}$)
       &  $\mu_{\scriptscriptstyle 5}$ 
       &  $< 1.0\times 10^{-5}$ \\[0in]
       $\bullet$ $U_{br}^{e\mu}$ \hspace*{.2in} ($e$-$\mu$ universality)
      &  $\mu_{\scriptscriptstyle 1}^2-\mu_{\scriptscriptstyle 2}^2$ 
       &  $(0.596 \pm 4.37)\times 10^{-3}$ \\[0in]
       $\bullet$ $U_{br}^{e\tau}$ \hspace*{.2in} ($e$-$\tau$ universality)
       &  $\mu_{\scriptscriptstyle 1}^2-\mu_{\scriptscriptstyle 3}^2$ 
       &  $(0.955 \pm 4.98)\times 10^{-3}$ \\[0in]
       $\bullet$  $U_{br}^{\mu\tau}$ \hspace*{.2in} ($\mu$-$\tau$ universality)
       &  $\mu_{\scriptscriptstyle 2}^2-\mu_{\scriptscriptstyle 3}^2$ 
       &  $(1.55 \pm 5.60)\times 10^{-3}$ \\[0in]
      $\bullet$ $\Delta{\cal A}_{e\mu}$ \hspace*{.1in} ($e$-$\mu$ L-R asymmetry)\hspace*{.5in} 
      &  $\mu_{\scriptscriptstyle 1}^2-\mu_{\scriptscriptstyle 2}^2$ + Rt. contrib. 
      &  $(0.346\pm 2.54)\times 10^{-2}$ \\[0in]
       $\bullet$ $\Delta{\cal A}_{\tau e}$ \hspace*{.1in} ($\tau$-$e$ L-R asymmetry)
       &  $\mu_{\scriptscriptstyle 3}^2-\mu_{\scriptscriptstyle 1}^2$ + Rt. contrib.
       &  $0.0043\pm 0.104$ \\[0in]
       $\bullet$ $\Delta{\cal A}_{\tau\mu}$ \hspace*{.1in} ($\tau$-$\mu$ L-R asymmetry)
       &  $\mu_{\scriptscriptstyle 3}^2-\mu_{\scriptscriptstyle 2}^2$ + Rt. contrib.
       &  $0.082\pm 0.25$ \\[0in]
       $\bullet$ ${\Gamma}_{\!\scriptscriptstyle Z}$ \hspace*{.3in} (total $Z^0$-width)
       &  $\mu_{\scriptscriptstyle 5}$
       &  $2.4948\pm.0075 \, \hbox{GeV}$ \\[0in]
       $\bullet$ ${\Gamma}_{\!\scriptscriptstyle Z}^{\scriptscriptstyle i\!n\!v}$\hspace*{.3in} (*)
       &  $\mu_{\scriptscriptstyle 5}$ 
       &  $500.1\pm5.4\, \hbox{MeV}$ \\[0in]
        $\bullet$ $Br$($Z^0 \to \chi^{0}_i\chi^{0}_j,
                                               \chi^{0}_j\nu) ; \; j \ne 1$
       &  $\mu_{\scriptscriptstyle 5}$ 
       &  $< 1.0\times 10^{-5}$ \\[0in]
      & & \\[-.1in]
       \framebox{$W^{\pm}$-coupling:}
        & & \\[.1in]
       $\bullet$        $\overline{\Gamma}^{\mu e}$ \hspace*{.2in} ($\mu \to e \nu \nu$)
       &  $m_{\nu_{\scriptscriptstyle 3}}\,/\,\mu_i$ ratio  &
        $0.983\pm0.111$ \\[0in]
       $\bullet$        $\overline{\Gamma}^{\tau e}$ \hspace*{.2in} ($\tau \to e \nu \nu$)
       &  $m_{\nu_{\scriptscriptstyle 3}}\,/\,\mu_i$ ratio  &
        $0.979\pm0.111$ \\[0in]
       $\bullet$        $\overline{\Gamma}^{\tau \mu}$ \hspace*{.2in} ($\tau \to \mu \nu \nu$)
       &  $m_{\nu_{\scriptscriptstyle 3}}\,/\,\mu_i$ ratio  &
        $0.954\pm0.108$ \\[0in]
       $\bullet$        $R^{\pi e}_{\pi \mu}$ \hspace*{.2in} ($\pi$ decays)
       &  $m_{\nu_{\scriptscriptstyle 3}}\,/\,\frac{\mu_{\scriptscriptstyle 1}}{\mu_{\scriptscriptstyle 5}}$ and $\frac{\mu_{\scriptscriptstyle 2}}{\mu_{\scriptscriptstyle 5}}$   &
        $(1.230\pm0.012)\times 10^{-4}$ \\[0in]
           $\bullet$        $R^{\tau e}_{\tau \mu}$ \hspace*{.2in} ($\tau$ decays)
       &  $m_{\nu_{\scriptscriptstyle 3}}\,/\,\mu_i$ ratio   &
        $1.0265\pm0.0222$ \\[0in]
        $\bullet$        $R^{\mu e}_{\tau e}$ \hspace*{.2in} (decays to $e$'s)
       &  $m_{\nu_{\scriptscriptstyle 3}}\,/\,\mu_i$ ratio &
        $1.0038\pm0.0219$ \\[0in]
       $\bullet$ $m_{\nu_{\scriptscriptstyle 3}}
|\tilde{B}^{\scriptscriptstyle L}_{e\nu_{\scriptscriptstyle 3}}|^2$ \hspace*{.2in} [$(\zb\zb)_{0\nu}$]
       &  $m_{\nu_{\scriptscriptstyle 3}}\,/\,\frac{\mu_{\scriptscriptstyle 1}}{\mu_{\scriptscriptstyle 5}}$  &
        $< 0.46\, \hbox{eV}$  (only for $m_{\nu_{\scriptscriptstyle 3}}\!<\!10\,\mbox{MeV}$) \\[0in]
        $\bullet$ BEBC expt.& $m_{\nu_{\scriptscriptstyle 3}}\,/\,\frac{\mu_{\scriptscriptstyle 1}}{\mu_{\scriptscriptstyle 5}}$ and $\frac{\mu_{\scriptscriptstyle 2}}{\mu_{\scriptscriptstyle 5}}$
       & \\[0in]
      & & \\[-.1in]
       \framebox{mass constraints:}
       & & \\[.1in]
        $\bullet$ $\nu_{\scriptscriptstyle 3}$ mass
       &  $\mu_{\scriptscriptstyle 3}$ 
       &  $<18.2\, \hbox{MeV}$ if $\nu_{\scriptscriptstyle 3} = \nu_{\tau}$ 
       \\[0in]
      &  $\mu_{\scriptscriptstyle 5}$  
    &  $<149\, \hbox{MeV}$ if $\nu_{\scriptscriptstyle 3} \ne \nu_{\tau}$ \\
     $\bullet$ $\chi^\pm$ mass
         &  $\mu_{\scriptscriptstyle 5}$ & $ > 70\,\mbox{GeV}$\\[.1in]
\hline\hline
\end{tabular*}
\end{table*}

A summary of all the  phenomenological processes and corresponding 
experimental bounds we have investigated is given in Table 1, details
of which we refer to our paper\cite{II}. Here we will only outline the major
features. The first group of these constraints  includes the three modified 
partial widths of $Z^0 \to \ell^+\ell^-$, in terms of their universality violation, 
their left-right asymmetries, and now nonvanishing off-diagonal
$Z^0\ell_i\ell_j$ couplings with the resulted dacays of $\mu$ and $\tau$
into three $\ell$'s. Among them the $\mu \to 3e$ bound represents a
particularly stringent constraint on the magnitude of the product
$\mu_{\scriptscriptstyle 1}  \mu_{\scriptscriptstyle 2}$.
The second group of constraints is from the charged current interactions.
Among them $R^{\pi e}_{\pi\mu}$ and $R^{\mu e}_{\tau e}$ are particularly
important. The two refer to the ratio of $Br(\pi \to e \nu\nu)$ to
$Br(\pi \to \mu \nu\nu)$ and that of $Br(\mu \to e \nu\nu)$ to
$Br(\tau \to e \nu\nu)$ respectively. Depending on the ratio among the 
$\mu_i$'s, a constraint from either of the two quantities typically bounds
the $\mu_{\scriptscriptstyle 5}$ value a bit below that admissible by
the $149\,\mbox{MeV}$ neutrino mass bound. Related constraints from
neutrinoless double beta decay and the BEBC beam dump 
experiment\cite{BEBC} also have important roles to play. The former, 
when applicable, imposes the most stringent constraint on  
$\mu_{\scriptscriptstyle 1}$. All the constraints taken together seem to prefer
a strong hierarchy --- $\mu_{\scriptscriptstyle 1} \ll 
\mu_{\scriptscriptstyle 2} \ll \mu_{\scriptscriptstyle 3}$.

The discussion of the tree-level neutrino mass above also serves to illustrate
our approach and the some major features of the result for all the other 
processes. The most interesting one is the very dramatic $\tan\!\zb$
dependence. The RPV effects on the leptonic sector are
suppressed by $\cos^2\!\!\zb$. The factor comes into all the nonstandard
parts of $Z^0\ell_i\ell_j$ couplings and into the charged current constraints
through the neutrino mass. The feature is easy to understand intuitively. The
properties of the leptons are changed as a result of their mixing,
through the $\mu_i$'s, with the fourth doublet $L_0$. However, the first order
effect of such a flavor mixing is unobservable in the mass eigenstates. The
observable effects of the mixing, and hence R-parity violation, comes in
as a result of the difference between the $L_i$'s and the $L_0$, which {it is}
really a higgsino --- it couples through the VEV of its scalar partner to
the gaugino. This is where the  factors of $\cos\!\zb$ come into the game.

\section{CONCLUSIONS}

We have presented the right framework for the phenomenological studies 
of the complete theory of supersymmetry without R-parity where 
no {\it a priori} assumptions needed to be imposed. Under the single-VEV
paramatrization framework, the trilinear RPV couplings, $\lambda$, 
$\lambda^{'}$, and $\lambda^{''}$, {\it do not} involve in tree level mass 
matrices. The full effect of any form of R-parity violation is 
characterized there by the three bilinear couplings $\mu_i$. The latter  
could give rise to a rich leptonic phenomenology even when analysis
is restricted only to the tree-level. We cannot do much more than giving
an outline of the full result here. Interested readers are referred to our detailed
report on the topic\cite{II}.

Of particular relevance here, a nonzero $\mu_{\scriptscriptstyle 3}$ 
represents the gaugino and higgsino content of the physical $\tau$ lepton.
While nonzero $\mu_{\scriptscriptstyle 1}$ and 
$\mu_{\scriptscriptstyle 2}$ characterize the corresponding properties
of the   physical $e$ and $\mu$, the experimental constraints on their
admissible magnitude  are much stronger, particularly stringent in the case
of  $\mu_{\scriptscriptstyle 1}$. The
constraint on  $\mu_{\scriptscriptstyle 3}$ is quite weak in many cases,
particularly for intermediate to large values of  $\tan\!\zb$. In case 
$\mu_{\scriptscriptstyle 3}$ happens to be substantial, it would make the 
$\tau$  substantially different from $e$ and $\mu$, which is a
very interesting scenario.

We quote some illustrative numbers here, for $M_{\scriptscriptstyle 2}
=\mu_{\scriptscriptstyle 0}=200\,\mbox{GeV}$: at $\tan\!\zb=10$, 
$\mu_{\scriptscriptstyle 5}$, with a dominating
$\mu_{\scriptscriptstyle 3}$ and a minor $\mu_{\scriptscriptstyle 2}$ 
contribution, can go to $101\,\mbox{GeV}$; 
at $\tan\!\zb=2$, $19\,\mbox{GeV}$; both giving 
$m_{\nu_{\scriptscriptstyle 3}}$ around  $110\,\mbox{MeV}$ and
$Br(\tau \to \mu e e) \sim 10^{-10}$, with deviations of $Z^0$ to $\tau$ width
and $\tau$ L-R asymmetry at $>1\%$ level for the latter case.

\end{document}